\documentclass{article}    

\usepackage{graphicx}

\title{\textbf{Quantum Brain States}}  
\author{Richard Mould\footnote{Department of Physics and Astronomy, State University of New York, Stony Brook,
\mbox{New York} 11794-3800; http://nuclear.physics.sunysb.edu/ \~{}mould}}  
\date{}    
   
\begin{document}             

\maketitle              

\begin{abstract}

      	If conscious observers are to be included in the quantum mechanical universe, we need to find the rules that engage
observers with quantum mechanical systems.  The author has proposed five rules that are discovered by insisting on
empirical completeness; that is, by requiring the rules to draw empirical information from Schr\"{o}dinger's solutions
that is more complete than is currently possible with the (Born) probability interpretation.  I discard Born's
interpretation,  introducing probability solely through probability current.  These rules tell us something
about brains.  They require the existence of observer brain states that are neither conscious nor unconscious.  I call
them `ready' brain states because they are on stand-by, ready to become conscious the moment they are stochastically
chosen.  Two of the rules are selection rules involving ready brain states.  The place of these rules in a wider
theoretical context is discussed.

\end{abstract}

\section*{Introduction}

 	Physicists have been overwhelmingly successful when it comes to matching ensembles of data with the statistical
predictions of quantum mechanics.  But quantum mechanics does not know how to represent physical systems that contain
conscious sub-systems, and it cannot adequately portray conscious observations of `individual' processes.  If an
observer watches a detector during a period of time in which a single particle might be captured, his experience can be
summarized in two parts.  (i) He will first see the detector in its original state $D_0$ prior to a particle capture,
and (ii) if there is a capture, he will see the detector jump to an excited state $D_1$.  The trouble with quantum
mechanics is that it cannot predict, represent, or describe this simple two-part phenomenon.  This is because
conventional quantum mechanics does not tell us \emph{how conscious brain states fold into the methodology of the
theory}.  Quantum mechanics is an empirical science that does a thorough job of predicting ensemble behavior; however, it
should also be able to predict \emph{everything} about an individual experience that is consistent with `quantum
reality'.  We will show below that quantum theory does not do that when conscious states become involved.  To fulfill
this simple requirement, the empirical predictions of the theory must be extended beyond their present limited range.  

 	This has been done by the author in a series of four papers \cite{RM1, RM2, RM3, RM4} that propose five rules that
apply to a variety of interactions like the one described above.  These rules are not intended to be an interpretation
of quantum theory, or a pedagogy of quantum theory; but rather, they are descriptions of how conscious states become
engaged with physical systems.  As such, they are synthetic statements about Nature that are shown to improve on the
(current) connection between theory and observation.   More will be said below about how these rules
might relate to a wider theoretical understanding.  

The validity of these rules rests on several assumptions:

\vspace{.3cm}
\textbf{A.} Every object in the universe is a quantum mechanical system or subsystem. So atoms, black holes, and in
particular ``brains" are quantum mechanical objects.  This includes conscious brains no less than unconscious brains.    

\textbf{B.} In the interaction to be studied, we assume that every component of the detector interacts with a separate
component of the (conscious or not) brain of the observer.  Therefore, brain variables become entangled with detector
variables.

\textbf{C.} All of the quantum mechanical uncertainties that are a property of an ensemble are also a property of each
member of the ensemble.  This means that the wave function $\psi$ of a particle describes the probability properties
of an individual particle, as well as the statistical behavior of an ensemble of those particles.  
\vspace{.3cm}

 It will be shown in Part I of this paper that these assumptions, plus a requirement that (i) and (ii) are predicted by
the proposed rules, imply that there exists a brain state that is neither conscious nor unconscious.  I call it a
\emph{ready} brain state.  There is no classical analog.  However, in the context considered, it will be clearly
demonstrated that \emph{if conscious brains exist, then ready brains exist}.  The properties of a ready brain state are
described in \mbox{Part I}, and are spelled out more completely in refs.\ 1-4.

In the above example, the observer has one experience before capture and another experience after capture. 
Corresponding to this, the proposed rules give us one solution prior to capture, and another solution after capture.  In
order to achieve this ``dual" solution, our rules must make a statement about the probability of capture that is
qualified by time.  This is accomplished by shifting the notion of probability from that of `square modulus', to that of
the probability per unit time -- or \emph{probability current}.  In this treatment, square modulus has no intrinsic
physical significance, so the Born interpretation is abandoned.  \emph{Only current flow is now considered to be
physically meaningful}.  This is not inconsistent with the way that the Born probability interpretation has been
successfully applied as will be demonstrated in Part I.  

According to von Neumann, state reduction in quantum mechanics means an irreversible change that occurs during a
measurement \cite{vN}.  This corresponds to a change that carries a pure quantum mechanical state into a mixture.  It
therefore replaces one kind of ensemble with another kind of ensemble.  Measurement in this sense is now thought to
occur the moment two components of a subsystem are sufficiently distinguishable that their correlated environments are
orthogonal to one another.  The cross term is therefore zero when the environmental variables are integrated out.  This
means that the components of the subsystem are \emph{locally incoherent}, so the object in its various possible states
takes on classical statistical properties.  Joos and Zeh call the resulting subsystem an \emph{improper mixture} \cite{JZ},
and the process is referred to as \emph{environmental decoherence} \cite{DG, TK, CW, JF}.  For a measurement of this
kind, no conscious observer needs to be present; and indeed, nothing happens during this measurement that is external to
the Schr\"{o}dinger process. This means that the universe will remain a superposition that is undisturbed by decoherence. 
It is a physicist who chooses the subsystem and decides to trace out its environment.  Nature does not eliminate the
environment or any other part of the wave function during this so-called measurement, and it does not destroy or in any
way alter the superposition of the total system.  Therefore, there is no objective ``collapse" of the wave function
associated with decoherence. 

It may still be meaningful to speak of an observerless measurement through decoherence, but the important question here
is: What happens when a conscious observer \emph{is} present?  It is my claim that a conscious observer whose brain
states are part of a larger system will interact uniquely with the system; and the resulting measurement will bring about
a genuine `collapse' of the wave function that is described in refs.\ 1-4 and Part I of this paper.  This collapse is
like the \textbf{R}-process of Penrose, but is different in ways that will become apparent as we proceed.

\section*{Motivation}

As stated above, one motivation behind this effort has been to find the rules that engage conscious states with 
quantum mechanical systems, and to fully describe an observer's conscious experience in individual cases.  There is
another motivation.  For 400 years physics has confined its attention to the `objective' world, refraining from any
serious consideration of the corresponding `subjective' world.   I believe that will change.  I believe that we will see
the development of theoretical tools that will enable physicists to reach across the Cartesian divide, stitching it
together with a proper mind/body theory.  The rules that I propose are \emph{not} such a theory.  They are only a set of
ad hoc rules that bridge the gap, for they lack the elegance of a broad based theory.  However, I believe that these
rules, or rules like them, will be part of the stitching in any correct mind-body theoretical framework.  They will
relate to that theory in the same way that the Balmer series relates to atomic theory.  Like any empirical formula by
itself, these rules are only a step; but I believe they are a step in the direction of a better theoretical understanding.

	There is still another motivation that is actually the original motivation.  Before I attempted the write these rules I
was convinced of the correctness of the \emph{parallel principle}, and (what I now call) rule (5) which implements that
principle in a special case.  This will be explained in Part II of this paper; but briefly, \mbox{rule (5)} establishes
a physical basis for the evolution of the so-called \emph{psycho-physical parallelism}, which refers to the way that
subjective reality parallels or mirrors objective reality.  Whatever may come of these rules, I believe that the
parallel principle is the key to making inroads into any trans-Cartesian theory.

\vspace {.5cm}
\begin{center}
\textbf{\Large{PART I}}
\end{center}

\begin{quote}  
Part I of this paper is a summary of the findings in refs.\ 1 and 2.  In these two papers, four of the five proposed
rules are formulated and applied in different gedanken experimental situations.    
\end{quote}

\section*{Interaction}

In the absence of an observer, the state of the above described particle/detector system is given by $\Phi(t) =
exp(-iHt)\psi_iD_i$, where $\psi_i$ and $D_i$ are the initial states of the incoming particle and the detector, and $H$
is the Hamiltonian divided by $\hbar$ .  Because $H$ includes an interaction term between $\psi$ and $D$, the resulting
state $\Phi(t)$ is an entanglement in which the particle variables and the detector variables are not separable. 
However, $\Phi(t)$ is a superposition (in the representation considered) so we can group its components in any way that
we like.  I form two major components: (1) those detector states $D_0$ in which there is no particle capture, and (2)
those detector states $D_1(t)$ in which there is a particle capture.  Let $\psi(t)$ represent the free particle as a
function of time, including all the incoming and scattering components.   It is therefore entangled with $D_0$.  But
since the detector is macroscopic, we may approximate $D_0$ to be a constant that can be factored out of $\psi(t)$.  The
captured particle is included in the component $D_1(t)$.  This gives

\begin{equation}
\Phi(t \ge t_0) = \psi(t)D_0 + D_1(t)
\end{equation}
where $D_1(t)$ is equal to zero at $t_0$ and increases in time, whereas $\psi(t)D_0$ is normalized to one at $t_0$ and
decreases in time. 

	Equation 1 ignores the orthogonal environmental terms that are attached to each component.  These are responsible for
decoherence that causes the components to become incoherent almost as soon as the second one comes into existence.   The
reader is reminded that the superposition in eq.\ 1 is not destroyed by this process.  It is preserved because it
represents the state of the entire universe, where the (not shown) environmental terms attached to each component have
\emph{not} been eliminated.  

It is here that I depart from the Penrose gravitational approach that would couple his \textbf{R}-collapse of the
superposition to the onset of decoherence between these two very different detector states \cite{RP}.  This collapse
would occur almost as soon as the interaction begins, so the  parameters of the interaction would have no
opportunity to influence the time at which the particle is captured.  This cannot be correct, for it would short circuit
the normal Hamiltonian dynamics of the interaction.  

	Contrary to this, I say that the superposition in eq.\ 1 will last as long as is necessary for the interaction to run
its course.  The timing of a collapse \mbox{(i.e., a particle capture)} is determined by the probability current
flowing from the first to the second component, and that is determined by the rate at which the particle wave approaches
and overlaps the detector at each moment.  For low-level radioactive emissions yielding particle waves that are spread
widely over space, the interaction in eq.\ 1 might last for hours or days.  In that case, the superposition will last
for hours or days.  So the `time' of a particle capture in this treatment is not a function of decoherence or
gravitational thresholds.  It is, \emph{and should be}, a function of the parameters of the incoming particle and the
cross section of the particle with the detector\cite{RM5}.

\section*{A Conscious Observer}

If an observer interacts with the detector during the above interaction, then assumptions \textbf{A} and \textbf{B}
tell us that two different components of the brain will be involved.  One of these ($B_0$) will be entangled with $D_0$,
and the other ($B_1$) will be entangled with $D_1$.  Equation 1 then becomes

\begin{displaymath}
\Phi(t \ge t_0) = \psi(t)D_0B_0 + D_1(t)B_1
\end{displaymath}
where again, $D_1(t)B_1$ is equal to zero at $t_0$.  

Penrose says that brain superpositions like these cannot exist for any perceptible length of time without an ``objective
reduction" \mbox{(see ref.\ 11, pp.\ 342,\ 355)}.  But the argument that we applied to the above detector applies equally
to a macroscopic brain.  The superposition in this equation cannot be immediately reduced because that would abort the
normal Hamiltonian dynamics.  The timing of the collapse (i.e., the particle capture) should be determined by the rate at
which the particle approaches and engages the detector/brain subsystem, not by the decoherence or gravitational properties
of the brain.  

For these reasons we dismiss the Penrose theory, but now there is another problem.  The above equation is \emph{as far as
the Schr\"{o}dinger process can go}.  It is a theoretical dead-end that does not portray the dual experience of the
observer.  It does not tell us which brain state is conscious, or exclude the possibility that they might both be
simultaneously conscious.  The equation is also reminiscent of the Schr\"{o}dinger cat paradox that suggests that there
are competing probabilities between $D_0$ and $D_1$ for the conscious attention of the observer.  And finally, it
encourages the belief that quantum mechanics can only be applied to ensembles.  But if quantum mechanics is to be a
complete empirical science (consistent with quantum reality), then it must fully describe the dual experience of this
observer, clarify the above ambiguities as to which states are conscious, and it must do so in individual cases.  

The solutions to Schr\"{o}dinger's equations of motion are normally connected to laboratory data through a single rule:
The Born Rule (i.e., the Born interpretation).  However, this rule is incomplete as well as technically wrong as will be
shown.  It does not \emph{ground} the Schr\"{o}dinger solution as fully as possible to laboratory observations.  We
therefore need to replace the Born rule by other rule(s) that make more complete connections.  These are the rules
proposed in this paper.  

We proceed by anticipating a satisfactory resolution of the above difficulties.  The desired result should include two
solutions: one that applies before particle capture, and one that applies after (possible) particle capture.  Particle
capture is occasioned by the system making a \emph{stochastic choice}, and the time that that happens is given by
$t_{sc}$.  The two solutions will then take the form: \mbox{$\Phi(t_{sc} > t \ge t_0)$} and $\Phi(t \ge t_{sc} > t_0)$,
where the first of these is

\begin{equation}
\Phi(t_{sc} > t \ge t_0) = \psi(t)D_0\underline{B}_0 + D_1(t)B_1
\end{equation}
The underline means that $\underline{B}_0$ is conscious, and the lack of underline under $B_1$ means that it is not
conscious.  The observational significance of this equation is clear.  It describes the experience of the observer
\emph{prior} to capture.  

The existence of the second component in eq.\ 2 may seem to introduce an ambiguity, but there is no empirical ambiguity
because the observer is unaware of it.  The second component is only on standby, ready to take over if and when the
particle is captured by the detector.  It is the ``ready" brain state referred to above.  It is physiologically capable
of consciousness, but only when it is stochastically chosen.  Until then, it is neither conscious nor unconscious, and
is unique to a quantum mechanical superposition.  The moment we distinguish $B_0$ from $B_1$ by saying that one is
conscious and the other is not, we confirm the existence of such a state.  This is what I meant by saying, ``if
conscious brains exist, then ready brains exist"\footnote{Our rules (3) \& (4) will dispose of the second component of
eq.\ 2 if a second observer observes the detector, or if the attention of the primary observer drifts in any way.}. 

Schr\"{o}dinger's equation applies continuously in eq.\ 2, increasing the square modulus of the second component at the
expense of the first.  This decrease in amplitude of the first component does not affect the `quality' of the conscious
experience of the observer associated with $\underline{B}_0$.  Subjective intensity, for example, is not diminished
during this decrease of square modulus as one would expect from classical physics.  Generally speaking, the qualitative
properties of a quantum mechanical state are determined by the intrinsic parameters of the state, and have nothing to do
with its amplitude.  So an observerÕs awareness of a conscious state like $\underline{B}_0$ is either on or it is off. 
It is on if the square modulus is positive, and off if it is zero.

\section*{Particle Capture --- A Stochastic Hit}

Probability current flowing from the first to the second component in eq.\ 2 represents the probability per unit time
that there will be a stochastic hit on the second component, and that corresponds to a particle being captured by the
detector.  When that happens, our rule (3) -- given in the next section -- will provide that the second component
becomes consciousness 

\begin{displaymath}
\psi(t)D_0\underline{B}_0 + D_1(t)B_1 \longrightarrow \psi(t)D_0\underline{B}_0 + D_1(t)\underline{B}_1
\end{displaymath}
and that simultaneously the first component goes to zero.  Therefore, the stochastic hit produces a second equation
(following directly from eq.\ 2) given by
\begin{equation}
\Phi(t \ge t_{sc} > t_0) =  D_1\underline{B}_1
\end{equation}
Here again the situation is clear: The observer is consciously aware of the detector in its `capture' mode.  Equations 2
and 3 are the dual solutions that our adopted rules draw from the dead-end of the Schr\"{o}dinger solution by itself.

\section*{The First Three Rules}

Equations 2 and 3 follow from the first three of our rules.  The first provides for the existence of a stochastic
trigger, the second provides for the existence of ready brain states, and the third provides for the collapse of the
state function. 
 
\vspace{0.3 cm}
\textbf{Rule (1)}:\emph{ For any subsystem of n components in an isolated system with a square modulus equal to s, the
probability per unit time of a stochastic choice of one of those components at time t is given by $(\Sigma_nJ_n)/s$,
where the net probability current $J_n$ going into the nth component at that time is positive\footnote{An isolated
system is thought of as not interacting with the rest of the universe.  If one wants to determine if the two components
in eq.\ 1 are locally coherent or incoherent, then the effect of the environmental terms must be considered.  But if one
only wants to find the probability per unit time of a stochastic hit on either component, it is sufficient to treat them
as an isolated system (i.e., ignoring the environmental terms).  Of course, the system will also be isolated if the
environmental terms are included, so the probability is not affected by the decision to include or not to include the
environmental terms.  An isolated system is part of the environment of another isolated system. }.}
 
\textbf{Rule (2)}: \emph{If the Hamiltonian gives rise to new components that are not classically continuous with the old
components or with each other, then all active brain states that are included in the new components will be ready brain
states.}  [note: an ``active" brain state is either conscious or ready.] 

\textbf{Rule (3)}: \emph{If a component that is entangled with a ready brain state B is stochastically chosen, then B will
become conscious, and all other components will be immediately reduced to zero.}
\vspace{0.3 cm}

Rule (1) insures that there will be a stochastic hit on the second component of eq.\ 2.  The basis representation of
the reduction is provided through \mbox{rule (2)} which insures that the second component in eq.\ 2 is a ready brain
state. \mbox{Rule (3)} converts eq.\ 2 to eq.\ 3.  Therefore, these rules go directly and uniquely from the solution of
Schr\"{o}dinger's equation to the empirical predictions that lie beyond that solution by itself; that is, to the dual
phenomenon represented by \mbox{eqs.\ 2 and 3}.  The rules supplement the Schr\"{o}dinger solutions by replacing the
Born rule. This is necessary if the theory is to give a full account of the `quantum' reality of observers.

I am not concerned with the question of \emph{why} these rules work, but only that they do work.  They work in the
simple sense that they include the observer in the quantum mechanical universe, and they ground the Schr\"{o}dinger
solutions to laboratory observations as completely as possible.  Empirical accountability is the only constraint that I
impose in the above example and in all of the other examples considered in refs.\ 1 and 2.  This severely narrows the
possibilities and therefore defines the rules rather sharply.  Understanding them will have to wait until there is a
proper theory of brain states and/or a more general trans-Cartesian theory.  I believe that these rules will naturally
emerge out of such a theory when it is found.  For instance, rule (2) is a selection rule that forbids the brainÕs
Hamiltonian from producing a discrete conscious brain state.  I assume that a correct theory of brains will put this
rule into an understandable context, and that rule (3) will also seem less ad hoc with the development of the wider
theoretical framework.

\section*{Everett's Many Worlds}

	The above rules forbid Everett's many-world interpretation of quantum mechanics.  Rule \#2 does not allow discrete
conscious states to be created except by stochastic choice; and when that occurs, all of the other components go to
zero.  Therefore, this treatment does not allow the consciousness of an observer to exist simultaneously in more than
one component of a superposition.  This negates Everett's claim that a conscious observer can coexist with other
conscious versions of himself as `they' evolve together along separate branches of the universal state function.  The
observer in each branch is said to have no direct knowledge of the other versions of himself.   

Everett's conjecture is the result of a theory of physics that is not properly grounded in experience.  It seems
plausible only because the Born rule by itself does not tie Schr\"{o}dinger's solutions to laboratory observations as
completely as it should.  This leaves room for the imaginative and unwarranted wanderings that take us into the many
unobservable worlds of Everett.

\section*{A Terminal Observation}

In a typical laboratory experiment, the observer interacts with the detector only after the interaction is complete; and
at that time, the equation of state is generally given a Born interpretation.  That is, the square modulus of each
component is said to equal the probability of finding the state in that component.  But we say in this treatment that
probability enters only through probability current, so the Born rule cannot be directly employed.  However, when rules
(1-3) are applied to a terminal observation, it is shown below that the Born rule is upheld, even though it is not
itself fundamental. 
  
When the interaction is complete at a time $t_f$, and before the observer has looked at the detector, the system in
eq.\ 1 will stabilize to 
\begin{equation}
\Phi(t > t_f) = \{\psi(t)D_0 + D_1(t_f)\}X
\end{equation}
where $X$ is the unknown brain state of the observer prior to his interacting with the detector.  At the moment $t_{ob}$
of observation, rule (2) requires that eq.\ 4 becomes
\begin{eqnarray}
\Phi(t \ge t_{ob} > t_f) &=& \psi(t)D_0X + D_1(t_f)X\\
&+& \psi'(t)D_0B_0 + D_1'(t_f)B_1\nonumber
\end{eqnarray}
where the primed components in the second row are zero at $t_{ob}$, and ready brain states $B_0$ and $B_1$ are entangled
with detector states $D_0$ and $D_1$ in accordance with  assumption \textbf{B}.  There are as yet no new conscious
states.  

The interaction in eq.\ 5 is entirely physiological, where probability current $J$ flows from the components in the
first row to components in the second row.

$$J_0 < 0 \hspace{.5cm} \mbox{and} \hspace{.5cm}J_1 < 0 \hspace{1.5cm} J_0' > 0  \hspace{.5cm} \mbox{and}  \hspace{.5cm}
J_1'> 0$$

So far, the interaction is determined by standard quantum mechanics, plus assumptions \textbf{A} and \textbf{B} and rule
(2).    It is at this point that we apply \mbox{rules (1 \& 3).} Rule (1) assures us that there will be a stochastic hit
on one of the two primed components during the physiological response.  Setting $n = 2$,

$$\int\!dt[J'_0 + J'_1] = \mbox{total probability of a hit} = 1$$
where currents $J'_0$ and $J'_1$ flow until the physiological interaction is complete.  

When one of the primed components is stochastically chosen, rule (3) causes the affected ready state to become
conscious and drives the other component to zero.  This reflects the laboratory experience in two parts: (1) the third
component in eq.\ 5 is chosen with a probability equal to  $\int\!dtJ_0'$, and (2) the fourth component is chosen with a
probability equal to  $\int\!dtJ_1'$.  The probability of one of the components being chosen is therefore the same as
that given by the Born rule.  However, that rule is replaced by one that introduces probability \emph{only} when current
flows from the system in the first row of  eq.\ 5 to the observer/system in the second row of eq.\ 5.   

	Our rules now adequately describe the observerÕs experience when he interacts with the detector before the beginning of
the particle/detector interaction, as well as after that interaction is complete.  It is shown in ref.\ 1 that the rules
also work if the observer enters the picture \emph{during} the particle/detector interaction. So in all of the ways that
an observer can become involved with the system, the rules pull additional (and correct) empirical information from the
theory.

\section*{Boundary Conditions}
   
Rule (3) provides for a `collapse' of the wave function when an observer is present.  It requires the \emph{elimination}
of components.  This is not an ensemble-to-ensemble reduction; but rather, it is a change that is discontinuously
inflicted on an individual system in which one or more components are made to disappear.  I call it a \emph{boundary
reduction}, because is has the effect of discontinuously changing the boundary conditions of the solution to
Schr\"{o}dinger's equation.  The Hamiltonian does not undergo change during this reduction, but the solution on which it
operates undergoes change.      

The boundary conditions of the universe emerging from the big bang are indeterminate with respect to position or
direction.  Throughout the inflationary period there are no concrete markers that distinguish one part of the universe
from another.  A single field supposedly dominated the inflationary expansion, and that is said to guarantee the
thermodynamic sameness of the horizon of the universe in every direction.  This is the point of the `horizon'
argument.  

	So the question is: If the universe was originally so featureless, why is there now so much differentiated structure? 
Suppose that the wave function of an oxygen atom is distributed (more or less) uniformly over a volume; and
simultaneously, that the wave function of a carbon atom is also distributed (more or less) uniformly over the same
volume.  Then in sufficient time, one would find the wave function of a carbon monoxide molecule that is distributed
(more or less) uniformly over that volume.  In the same way, the (more or less) uniformly distributed GUT particles
that emerged from the inflationary expansion will combine in such a way that -- in sufficient time -- one would find
wave functions of galaxies that are (more or less) uniformly distributed throughout the universe.  There is clearly
something wrong with this scenario.  The contemporary universe is not quantum mechanically amorphous to this extent. 
This means that there must have been post-inflation boundary conditions imposed on the universe, and this could only
have happened through a mechanism that brought about a series of collapses of the original universal wave function.  The
question then is: How did that happen?  What was that mechanism?

	That mechanism must be external to the equation of motion, for differential equations cannot modify their own boundary
conditions (i.e., they cannot select one `eigen-galaxy' and discard another).  However, an external chooser (like our
stochastic trigger) can do that.  It can say: This galaxy stays, and that one goes.   

It is therefore necessary to add a choice mechanism of some kind to the differentially based formalisms of physics in
order to provide for post-inflationary boundary conditions.  A spontaneous reduction theory like that of Ghirardi,
Rimini, and Weber would qualify, as would a gravitational reduction theory like that of Penrose, whatever else may be
said for or against these theories.  Environmental decoherence does not qualify as a mechanism because it does not
choose between `eigen-objects'.  Of course, environmental arguments may be used in conjunction with other theories that
do qualify.      

Rule (3) qualifies, for it provides a boundary-defining stochastic trigger.  So the question is: Is the rule (3)
reduction a part of a more general reduction, or is it possible that rule (3) describes the \emph{only} example of a
boundary reduction in the universe?  Put differently: Is a ready brain so special that it is the only configuration of
matter that is capable of responding to a stochastic hit?

\section*{Objective Reduction: Rule (1a)}

It is not difficult to construct a rule that provides for the existence of boundary reductions in the absence of an
observer.  I will call it \emph{objective reduction} \mbox{rule (1a)}.

\vspace{0.5 cm}
\textbf{Rule (1a)}: \emph{If a component of a superposition is locally incoherent with other components in the
superposition, and if it is stochastically chosen, then all those other components will be immediately reduced to zero.}
\vspace{0.5 cm}

If rule (1a) were in play, then the first particle/detector component $\psi(t)D_o$ in eq.\ 1 would become zero the
moment the second component $D_1(t)$ is stochastically chosen.  In general, this rule would provide for an
`observerless' measurement (including a genuine boundary defining collapse) associated with environmental decoherence. 
In addition, this reduction would preserve Hamiltonian dynamics, so it could not be explained by the Penrose theory
without some modification.  Of course, I have not tried to `explain' any of these rules.  I have only sought rules that
give the right results, where explanations are left to the possibility of a wider theoretical framework.  The trouble
is, the observer cannot generally know if rule (1a) gives any result at all; for the effect of
the rule is experimentally indistinguishable from the effect of rules (1-3) by themselves.  

Rule (1a) applies in the absence of an observer, but the above rules (1-3) are still required when an observer is
present.  We need rule (1) in any case, to provide a stochastic trigger.  In addition, rule (2) is necessary to provide
for the existence of ready brain states during active observation, and rule (3) is needed to turn a stochastically
chosen ready brain state into a conscious state.  Rules (4) and (5) will also be found to apply in the presence of an
observer, and will be in play with or without rule (1a).  Therefore, \emph{when an observer is present, rule (1a) is
superfluous; and when an observer is not present, the effects of rule (1a) cannot be observed.}  The choice is to adopt
rule (1a) together with a modification of rule (3) that does not provide for a separate reduction [call it rule (3mod)],
\emph{or} to adopt the rules (1-3) as originally stated.  The observer cannot tell the difference\footnote{If an earlier
boundary condition is uniquely correlated with a later observation, and if the observation is stochastically chosen by
the rules (1-3) proposed above, then it will be impossible for the observer to know if the earlier boundary was also
chosen by an earlier application of rule (1a).  This will be true even if the boundary is not uniquely correlated with
the observation.  Causal correlations that connect back to a range of possible earlier `incoherent' boundaries will
result in uncertainty in the boundary conditions that led to the observation.  So the precise nature of the earlier
boundary cannot be verified by the observer, and there will be no reason for him to believe that any one of those
possible boundaries was (or was not) the \emph{only} boundary by virtue of an earlier rule (1a) reduction.  Therefore,
to this extent, the effects of rule (1a) are empirically indistinguishable from the effects of \mbox{rule (3)}. An
exception may be possible if causal connections go back to a range of earlier `coherent' boundaries; however, an example
like this is worked out in ref.\ 13 with results that also support indistinguishably.}.  In another paper \cite{RM6},
it is shown that the difference is unobservable in all of the examples that have been worked out in  refs.\ (1-2).

This choice of rule (1a) or rule (3) will be decided by determining which best fits into a wider theoretical
framework.  It is my belief that the missing link in all of this is a genuine trans-Cartesian connection.  The brain
\emph{is} special because it supports consciousness, and when we have finally understood this phenomenon it is my belief
that we will also have the theoretical framework that will favor the rules that exploit the uniqueness of a ready
brain state -- that is, the rules as originally given.  

This is the view put forward von Neumann, Wigner, London and Bauer.  It does not mean that consciousness `causes' state
reduction as is sometimes said; but rather, that there is a common cause of both.  Rule (3) says that a stochastic hit
on a ready brain state gives rise to both consciousness and a boundary reduction.   \emph{It} is the cause of both.

The remainder of this paper does not require the reader to accept this point of view.  Having stated my position, I
return to the primarily question concerning what happens when a conscious observer \emph{is} involved in an interaction --
not with what might happen otherwise \footnote{There is one consequence of this idea concerning the importance of an
observer that people find incredible.  It is that there may be significant parts of the universe that remain quantum
mechanically amorphous because they still haven't been observed by a conscious observer -- as for instance, large parts
of the surface of Venus.  I agree that this is incredible.  But it is no more so than many other things about quantum
mechanics -- as for instance, the diffraction properties of a single particle.}$^,$\footnote{I do not accept Penrose's
reduction criterion because I see no reason why a metrical superposition should not exist so long as all of its
components are mapped onto the same topological space.  However, I agree that the singularities associated with the
centers of black holes cannot be plausibly superimposed.  Therefore, the appearance of a black hole must have
established a boundary condition that uniquely located its center and excluded everything else from that location. 
This must be an exception to the `observer' requirement, for it means that the positions of most galaxies (and the
spaces between them) were probably determined long before the appearance of ready brains and rule (3) boundary
reductions.}$^,$\footnote{My rejection of Penrose's \textbf{R}-process should not reflect on the Penrose-Hameroff idea
that coherent superpositions within microtubules might have something to do with consciousness, or brain readiness. 
That may or may not be the case. }$^,$\footnote{If the environment of a system contains another system that undergoes a
state reduction by either rule (1a) or rule 3, then the square modulus $s$ of the entire system may change (i.e.,
decrease) in the middle of the primary interaction that is being considered.  This amounts to a mid-interaction change
in boundary conditions, and will be accompanied by a corresponding change (i.e., decrease) in the current flowing into
each component.  Since rule (1) involves the ratio of $J/s$, and applies at each moment of time, the new boundary will
not effect the probability per unit time.  This means that rule (1) will correctly apply with one value of $s$ before
the environmental reduction, and with another value of $s$ after the reduction.  In general, the square modulus of an
isolated system that includes environmental terms may decrease rapidly due to environmental state reductions.  Although
rule (1) is completely valid in this fluid situation, it will be easier to stabilize $s$ by treating (isolated) systems
in which the environmental terms have been excluded.}.

\section*{Two Observes -- Rule (4)}

We have seen that rules (1-3) work well in all of the ways that an observer might interact with the particle/detector
system in eq.\ 1.  These rules also work well in all of the ways that two observers might interact with the system -- with
one exception.  Rules (1-3) by themselves give rise to an anomaly that must be remedied.  To this end, a fourth
rule is added that puts an important restriction on ready brain states.  

Suppose the first observer is on board with the detector from the beginning as in eq.\ 2, and let the interaction
terminate without a capture.  The equation would then take the form 

\begin{displaymath}
\Phi(t > t_f) = \{\psi(t)D_0\underline{B}_0 +  D_1(t_f)B_1\}X
\end{displaymath}
where $t_f$ is the time that the particle/observer interaction ends, and $X$ represents the brain state of the second
observer who has not yet interacted with the system.  The first term in a product like $B_1X$ will refer to the first
observer, and the second term will refer to the second observer.  Current has stopped flowing at this stage, so
physically, it corresponds to the observer just staring at the detector with no hope that the particle can still be
detected -- since the interaction was over at $t_f$.  The second observer then looks at the detector at time $t_{ob(2)}$
which, according to rule (2), gives  

\begin{eqnarray}
\Phi(t \ge t_{ob(2)} > t_f) &=& \psi(t)D_0\underline{B}_0X + D_1(t_f)B_1X\\
&+& \psi'(t)D_0\underline{B}_0B_0 + D_1'(t_f)B_1B_1\nonumber
\end{eqnarray}
where current now flows into the primed components in the second row that are equal to zero at $t_{ob(2)}$.  If there is
a stochastic hit on the third component in \mbox{eq.\ 6}, the resulting state equation will immediately become  
$\psi'(t \ge t_{ob(2)})D_0\underline{B}_0\underline{B}_0$, which corresponds to both observers witnessing the detector
state $D_0$ that has been passed over without a particle capture.  However, if there is a stochastic hit on the fourth
component in eq.\ 6, the state equation will become $D_1'(t_f)\underline{B}_1\underline{B}_1$, according to which both
observers witness a particle capture.  But that is impossible, since the first observer can testify that the interaction
has already ended without a capture.  Therefore, the latter transition must be forbidden.  This is accomplished in a
general way by requiring another selection rule that  reflects the properties of the brain's Hamiltonian.

\vspace{0.5 cm}
\textbf{Rule (4)}: \emph{A transition between two components is forbidden if each is an entanglement containing a
ready brain state of the same observer.}
\vspace{0.5 cm}

There is an entangled ready brain state $B_1$ of the first observer appearing in both the second and the fourth components
of eq.\ 6.  Rule (4) therefore says that no current can flow between them, so the anomalous state
$D_1'(t_f)\underline{B}_1\underline{B}_1$ cannot be chosen.  It is shown in ref.\ 1 that the fourth component in eq.\ 6
will not even appear in that equation.  It is also shown in ref.\ 1 that rules (1-4) work perfectly well in all of the
ways that two observers might interact with the system and with each other -- and there are now no
exceptions\footnote{In eq.\ 6, the probability of a capture is found by applying rule (1) with $n = 1$ to the second
component prior to $t_f$.  The probability of no capture is found by applying rule (1) with $n = 1$ to the third
component after $t_{ob(2)}$.  The sum will equal 1.0.}.  

\begin{quote}
The rules so far do four notable things. (1) They impose boundary conditions that contribute to the structured universe of
our experience, and (2) they draw empirical content from quantum mechanics (e.g., the duel phenomenon) that is \emph{not}
predicted by the Schr\"{o}dinger theory plus the Born rule.  (3) They amend the Born rule to apply to probability current
rather than square modulus.  One consequence of this is that there is no need to insist on the normalization of system
states.   And finally, (4) these rules give conscious observers a way to engage quantum mechanical
systems.  In the process, they introduce non-classical ready brain states that have a crucial but limited role to play. 
These are brought into existence by the second rule, they are stochastically chosen by the third rule, and their influence
is limited by the fourth rule.  If nothing else, ready brains define the basis states of a boundary reduction.    
\end{quote}

\section*{Schr\"{o}dingerÕs Cat Experiment}

	No new rules or assumptions are introduced in ref.\ 2, where we turn attention to the famed Schr\"{o}dinger cat
experiment.  The cat presents us with a different operational situation that can be used to further test the rules and
assumptions that have already been adopted.  They all pass the test.   

	There are two versions of the Schr\"{o}dinger cat experiment.  In version I, a conscious cat interacts with a device
that splits it into two components.  In the first component the cat is made unconscious, and in the second it remains
conscious.  This is another case in which the equations of motion give rise to a single solution without resolving the
question of the consciousness or unconsciousness of the brain states that are involved.  When our rules (1-4) are
applied to this case, they again tease out two solutions: one corresponding to the case in which there is a stochastic
hit that leaves the cat unconscious, and one in which there is no stochastic hit, leaving the cat fully conscious.  

	The rules also resolve all issues that might arise when an outside observer is brought into the picture.  The observer
can be injected into the scene before there has been a stochastic choice, or during the operation of the mechanical
devise that renders the cat unconscious, or after the experiment is over.  In all of these cases the rules correctly
predict the conscious experience of the observer and the cat, leaving no ambiguity of the kind that typifies the
Schr\"{o}dinger solution by itself.  

In version II, the cat is initially unconscious and is made to interact with a device that splits it into two
components.  In the first component the cat is aroused to consciousness, and in the second it remains unconscious. 
Again we examine what happens when an outside observer interacts with the system at any time during the experiment.  In
all of these cases, rules (1-4) convert the Schr\"{o}dinger solution into a correct description of the conscious
experience of the observer and the cat.  Here again, the rules remove all of the ambiguity that normally plague quantum
mechanical solutions when conscious beings are involved.  

	Finally in ref.\ 2, the version II cat is assumed to have an internal alarm that competes with the external alarm, where
each is capable of arousing the cat.  Rules (1-4) correctly describe the catÕs various
experiences in these circumstances.  Again an external observer is allowed to intercede at any time during the
experiment; and again, the rules resolve all of the quantum mechanical paradoxes that are usually associated with
the Schr\"{o}dinger cat. 

	These rules work as well as they do in the cat experiments because they properly ground the Schr\"{o}dinger solutions
in experience, leaving no fuzzy ambiguities.  

\pagebreak

\vspace {.6cm}
\begin{center}
\textbf{\Large{PART II}}
\end{center}

\begin{quote}  
Part II is a summary of the findings in refs.\ 3 and 4.  In these two papers we take account of the fact that a conscious
brain state cannot be sharply defined, so it must have a pulse-like nature.      
\end{quote}

\section*{A Conscious Pulse}

References 1-2 make the unrealistic assumption that brain states have sharply defined physiological variables.  But any
real conscious state will span a range of physiological states, so refs 3-4 define the properties of a brain
\emph{pulse}.  

In ref. 3, rule (3) is amended to accommodate this change.  The amendment is called rule (3a), and it asserts that when
a conscious state is stochastically chosen it dissolves into a pulse of states that reflect the brains limits of
resolution.  This dissolution is affected by the brain's Hamiltonian.  In any final presentation of the rules, rule (3a)
will be joined with rule (3) to give a full account of what happens when there is a boundary reduction.  The other rules
are not affected by a change to pulses from sharply defined states.  

It is shown in ref.\ 3 that when an observer tries to distinguish between two apparatus eigenvalues that are too close to
be resolved by a conscious pulse, the superposition containing these two eigenvalues cannot be fully reduced by rules
(1-4).  The sharpness of the reduction depends on the extent to which the resolution curve overlaps the two
eigenvalues.  In general, the observation of a superposition results in another superposition, due to resolution
difficulties. 

	Reference 4 continues the discussion of the properties of a conscious pulse, but with the addition of the final rule
(5).  To explain this, we need to first introduce some ideas concerning the relationship of consciousness to
physiology.

\section*{The Parallel Principle}

	It is generally accepted that the subjective world of our personal experience corresponds in critical ways with the
objective world that exists outside of ourselves.  It is assumed that formal relationships can be found in the
subjective world that parallel or mirror the formal relationships that exist in the objective world -- and this is the
basis of epistemology in physics.  Von Neumann calls it the \emph{psycho-physical parallelism}.  Why a psycho-physical
parallelism should exist at all is open to question, inasmuch as these two separate realms of reality have such
different natures.  There is no obvious reason why one of these worlds should pay attention to the fortunes or
machinations of the other.  Responding to this point, Leibniz claimed that we are compelled to believe in a          
Pre-established Harmony between these realms that is arranged by God.

	Opposed to this, the \emph{parallel principle} says that the psycho-physical parallelism is the consequence of natural
evolutionary processes.  It claims that the conscious evolution of a species develops in parallel with the physiological
evolution of the species; and for this to happen, there must be an \emph{interaction} between consciousness and
physiology.  The general mechanism for this interaction is described in a previous paper \cite{RM7}.  

	The rules that have been considered so far allow one to believe that consciousness is only epiphenomenal: that is, it
may be regarded as an insubstantial by-product of physiology that has no counter-influence of its own.  It is possible
to believe that consciousness just tags along with the rule (3) reduction that is set in motion by the stochastic choice
in rule (1).  But for the parallel principle to work, consciousness must do more than tag along.  It must do something. 
It must have an influence on physiology.  Its presence must give an evolutionary preference to one kind of physiology
over another, and to achieve this we adopt rule (5).  

	The mechanisms for this influence are likely to be quite involved, so I narrowed down my considerations to just one
kind of consciousness -- \emph{pain consciousness}.  It is possible that the first conscious experience of a species was
`pain'; but for pain to be influential there must be a behavioral consequence when it appears.  If that behavior is
beneficial to the species in its evolutionary struggle, then a parallelism will be established between the appearance of
pain and a response that is beneficial to the species.  

	The logic of this argument may seem to be obvious and well known.  But the implications are dramatic if two things are
made clear.  First, the term `pain' is not a euphemism for just another physiological sub-routine.  Pain here refers to
the subjective experience itself, considered as something apart from physiology.  It that were not so, the parallel
principle would not work.  If all causal influences were basically physiological, then there is no reason why
subjective experience should in any way parallel or mirror the physical world.  Subjective impressions and images would
then lead their own irrelevant existence.  They might be stimulated by physiology, but their content would be
nonsensical.   Therefore, for there to be a disciplined relationship between physiology and consciousness, there must be
a meaningful interaction between the two.  Second, the term `pain' for one species may not have the same qualitative
significance as ÔpainÕ for another species.  The psycho-physical parallelism does not depend on this `quality', but
requires only that a certain kind of conscious experience is associated with a certain kind of behavior, and that this
combination is an evolutionary success.  Therefore, all we have to do is establish that a conscious experience of some
kind has a lawful influence on behavior, and evolution will do the rest.  The parallel principle will be satisfied.  

	 When I say `pain' in the following, I refer to an experience that arose somewhere in the evolutionary past of the
subject species, and became correlated with a successful behavioral response.  To achieve this kind of success, I claim
that pain itself (not a physiological substitute) must alter physiological brain states.  If there are no behavioral
consequences of this alteration, then the newly introduced phenomenon will be of no consequence, and will fade away
over time as a useless subjective appendage.  However, if there are behavioral consequences, then evolution will shape
those consequences in such a way as to validate the parallel principle.  

	The mechanism for this physiological influence is to be found in the structure of the conscious pulse.  It is assumed
that the primordial organism has the conscious experience in question (i.e., pain), and that this manifests itself in a
conscious pulse of finite width.  Reference 4 has a detailed discussion of how such a pulse might be formed, and how its
participating states will be spread over a range of greater or lesser pain.  

\begin{figure}[h]
\centering
\includegraphics[scale=0.8]{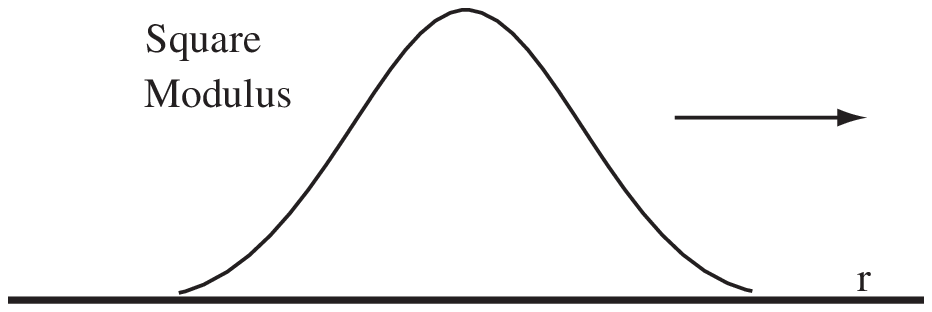}
\center{Conscious Pulse of Pain States}
\end{figure}

The accompanying figure shows that pulse as a function of a variable $r$, which (according to the formation mechanism
that I imagined) is proportional to the number of opiate receptors that are occupied by molecules that stimulate receptor
activity (see refs.\ 4, 14).  A stimulated opiate receptor supports euphoria and/or analgesia.  This means that the
leading edge of the pulse refers to conscious states that are less painful than the states in the trailing edge.  We
then introduce rule (5) 

\vspace{0.5 cm}
\textbf{Rule (5)}: \emph{If two states within a conscious pulse represent different degrees of pain, then the square
modulus of the state with lesser pain will increase at the expense of the state with the greater pain. }
\vspace{0.5 cm}

	Because the trailing edge states are the more painful of the two, rule (5) will cause the leading edge in the figure to
grow at the expense of the trailing edge.  As a result, the pulse will drift to the right as indicated by the arrow.  So
\mbox{rule (5)} brings about a migration of the brain's conscious pulse that will result in physiological changes that
may very well result in behavioral changes.  Again, see details in refs.\ 4 and 14.  The organism may therefore be
altered in a way that may in the long run be beneficial to the species.  If that happens, then it will be genetically
established that pain will be associated with a behavioral response that furthers the survival of the species.  This
will be the beginning of a psycho-physical parallelism, established in accordance with the parallel principle.  

\section*{Experimental Test}

	If rule (5) is correct, then there will be physical consequences that should be observable in principle.  A test of
rule (5) is proposed by the author in another paper where two  experiments are suggested: (1) a PET (Positron Emission
Tomography) scan of a human subject experiencing pain, and (2) an autoradiography of a rat experiencing pain \cite{RM8}. 
I will briefly describe the PET experiment.

	\emph{Agonists} molecules attach to opiate receptors and stimulate a response that is recognized by the subject as
euphoria and/or anesthesia.  \emph{Antagonist} molecules also  attach to these receptors, but their
only effect is to block the attachment of agonist molecules.  In this experiment, a subject experiencing pain is injected
with a certain ratio of opiate agonist to antagonist molecules, where one or the other is radioactively labeled.  The
extent of positron emission is then measured in parts of the brain where there are high densities of opiate receptors. 
There are four separate scans. 

\vspace{.2cm}
(a) the agonist is labeled and the dose is pharmacological.

(b) the antagonist is labeled and the dose is pharmacological.

(c) the agonist is labeled and the dose is subpharmacological.

(d) the antagonist is labeled and the dose is subpharmacological.
\vspace{.5cm}

Let $r$ be the ratio of positron emission between (a) and (b), where this is to be compared with the ratio $r$ of
emission activity between (c) and (d).  The variable $r$ in each case is equivalent to the number of receptors that are
occupied by agonist molecules, compared with the number of receptors that are occupied by antagonist molecules.  On
purely physiological grounds, the ratio ought to be the same in both cases (a/b and c/d) because the difference has only
to do with dose, and the doses are well below saturation.  However, the subject is conscious of the effects of the
injection in the first case (i.e., the pharmacological dose), but not in the second case (i.e., the subpharmacological
dose).  If rule (5) is correct, then the drift of the conscious pulse in the first case should alter the result by
increasing the number of agonist molecules relative to the number of antagonist molecules.  We should therefore measure

\begin{center}
$r$ (between a and b) $> r$ (between c and d)
\end{center}
thereby validating rule (5).   This validation depends on a change in $r$ that cannot be accounted for on purely
physiological grounds.   There is reason to believe that PET technology is up to this degree of accountability, or soon
will be.

\end{document}